\documentstyle[aps,prl,psfig,epsfig]{revtex}         % PRC-Layout
\newcommand \be{\begin{eqnarray}}
\newcommand \ee{\end{eqnarray}}

\begin{document}
\draft
\twocolumn[\hsize\textwidth\columnwidth\hsize
           \csname @twocolumnfalse\endcsname  
\title{Mesoscopic fluctuations of the Density of States and  
  Conductivity in the middle of the band of Disordered Lattices}
\author{E.P.Nakhmedov$^{1,2,3}$ \and V.N.Prigodin$^4$ and E.\c Sa\c
  s\i o\~glu$^3$}
\address{$^1$ Max-Planck Institut f\"ur Physik Komplexer Systeme,
  N\"othnitzer str.38, 01187 Dresden, Germany\\
$^2$Azerbaijan Academy of Sciences, Institute of Physics, H.Cavid 33,
Baku, Azerbaijan\\
$^3$ Fatih University, B\"uy\"uk\c cekmece, Istanbul, T\"urkiye\\
$^4$ Physics Department, The Ohio State University, Columbus, OH 43210
- 1106, USA}
\date{\today}
\maketitle
\begin{abstract}
The mesoscopic fluctuations of the Density of electronic States (DoS)
and of the conductivity of two- and three- dimensional lattices with randomly
distributed substitutional impurities are studied. Correlations of the
levels lying 
above (or below) the Fermi surface, in addition to the correlations of
the levels lying on opposite sides of the Fermi surface, take
place at half filling due to nesting. The Bragg reflections
mediate to increase static 
fluctuations of the conductivity in the middle of the band which
change the distribution function of the conductivity at half- filling.
\end{abstract}
\vskip2pc]

Recent studies of electronic level statistics in disordered
systems have shown  \cite{efetov,as,akl,prigodin,ag,kl,mirlin} that existence of repulsions between the
levels in metallic phase results in a realization of the
Wigner- Dyson statistics, \cite{mehta}. Energy levels in a sufficiently
dopped $d>2$ dimensional electron gas become uncorrelated in an insulator
phase and the level distribution obeys to Poisson statistics,
\cite{sss}. However, overlapping of one--particle states with
different energies leads to level correlations which change their
distribution.  

Our recent studies of weak localization effects in two--dimensional
(2D) square lattice and in three- dimensional (3D) cubic lattice with
substitutional impurities have revealed that Bragg reflections (BR)
due to commensurability of the electronic 
wavelength, $\lambda$, and a lattice spacing, $a$, strongly change the
localization picture, \cite{nko,nfok,np}. The DoS vanishes on
the Fermi surface for non--interacting electrons in a $2D$ lattice and
it acquires a small dip on the Fermi surface 
of $3D$ simple cubic lattice  with approaching half--filling,
\cite{nko,nfok}. Nevertheless, electron--electron (e-e) interactions 
give a positive  quantum correction to the DoS,\cite{np},  which
compensates the Altshuler--Aronov's negative logarithmic corrections to
the DoS of $2D$ systems,\cite{aa1,aal}. Therefore it is
interesting to clarify how static fluctuations of physical parameters
of $2D$ disordered lattice with nested Fermi surface  are changed by
BR as the half--filling is approached. 

In this paper we consider the effect of BR on the level
statistics and on conductivity fluctuations of $2D$ and $3D$
disordered lattices with a half- filled band.
A particular characteristics of level spectra is the two--level
correlation function,
\begin{equation}
R(\epsilon,{\epsilon}') =\frac{1}{{\rho}_{od}^2}\{\langle \rho
(\epsilon )\rho ({\epsilon} ')\rangle - 
\langle \rho (\epsilon )\rangle \langle \rho ({\epsilon}')\rangle \},
\end{equation}
where $\langle ...\rangle$ means averaging over impurity
realizations. $\rho_{od}$ is
the DoS of the $d$--dimensional ($d=2,3$) lattice calculated in the Born
approximation:
$\rho_{o2} = \frac{2}{(\pi a)^2 t}\ln({\epsilon}_F
min\{\tau_o,\frac{1}{|\epsilon |}\})$, and $\rho_{o3} = const - \frac{2}{{\pi}^2t^{3/2}a^3}
\sqrt{|\epsilon | -t}$, ( at $ |\epsilon |\approx t$) with $t$ and
$\tau_o$ being the tunneling integral for nearest--neighbor 
sites and the relaxation time for elastic impurity scattering,
respectively.

By using the formula for the DoS,
$\rho (\epsilon ) = \frac{1}{2\pi i}\int\frac{d{\bf r}}{v}\{G_A({\bf
  r},{\bf r'};\epsilon) - G_R({\bf r},{\bf r'};\epsilon ) \}$,
which relates $\rho (\epsilon )$ to the retarded $(G_R)$ and advanced
$(G_A)$ Green's functions, $R(\epsilon, {\epsilon}')$ can be expressed
as,
\begin{eqnarray}
&&R(\epsilon , {\epsilon}')= -\big(\frac{s}{2\pi v\rho_{od}}\big)^2\int
d{\bf r}\int d{\bf r'}\nonumber\\
&&\{\langle G_A({\bf r},{\bf r};\epsilon)G_A({\bf
  r'},{\bf r'};{\epsilon}')\rangle + \langle G_R({\bf r},{\bf r};\epsilon)G_R({\bf
  r'},{\bf r'};{\epsilon}')\rangle - \nonumber\\
&&- \langle G_R({\bf r},{\bf r};\epsilon)G_A({\bf
  r'},{\bf r'};{\epsilon}')\rangle - \langle G_A({\bf r},{\bf r};\epsilon)G_R({\bf
  r'},{\bf r'};{\epsilon}')\rangle \nonumber\\
&&-4 Re \langle G_R({\bf r},{\bf r};\epsilon)\rangle Re \langle G_R({\bf
  r'},{\bf r'};{\epsilon}')\rangle \},
\label{Rcor}
\end{eqnarray}
where  $v$ is the 'volume'  and $s$ is the factor of spin degeneracy.   
Far from half--filling the correlators $RA$ and $AR$ in Eq.(\ref{Rcor})
give only contributions to the two--level correlation
function,\cite{as}. However, existence of the electron--hole symmetry for
nested Fermi surfaces gives rise to considerable contributions of the
$RR$ and $AA$ correlators to $R(\epsilon,\epsilon')$ in Eq.(\ref{Rcor}).
The Fermi surface of a $d$-dimensional lattice with the energy spectrum of
$\epsilon ({\bf p})= t \sum_{i=1}^d [1 - \cos (p_ia)]$ becomes nested at
half- filling, when $\epsilon_F =dt$, which permits an electron--hole symmetry,
$\epsilon({\bf p}+{\bf Q}) - \epsilon_F = -[\epsilon({\bf p}) -
\epsilon_F]$, with respect to the nesting vectors ${\bf Q} = \{\pm
\pi/a,\pi/a\}$ for $2D$ and ${\bf Q} = \{\pm
\pi/a,\pm \pi/a,\pi/a\}$ for $3D$ lattices. New singular impurity
blocks take place at half--filling with particle--hole symmetry, which
are referred to as the $\pi$-Diffuson ($D_{\pi}$) and the
$\pi$-Cooperon ($C_{\pi}$). The $\pi$--Diffuson ($\pi$--Cooperon) has
a diffusion pole at large $\propto {\bf Q}$ momenta differences (total
momenta) and small total energies of the electron and the hole (of two
electrons), \cite{nko,nfok}. 

As it is known, an interference between the  self- intersecting
trajectories leads to the weak localization of an electronic wave
function, 
(see,\cite{aa1}). The electron passing the loop (e.g. in
Fig.1a), which is formed on the trajectory due
to multiple scattering on impurities with small tilted angles, in
clock- and counterclockwise, reduces its transmission
probability. However, for a nested Fermi surface each act of impurity
scattering is accompanied by BR 
with large ($\sim \pi$) scattering angle (see 
Fig.1b), which strongly changes the weak localization picture
of free electron gas. Worthwile to notice that the electron- hole
symmetry effects in the $2D$ case strongly differ from
that in the one-dimensional ($1D$) case, where BR act as a
destructive factor of localization and result in big
effects (like the Dyson singularity in the $1D$ DoS, \cite{dyson}) only
in the absence of
forward scattering (see,e.g.\cite{gd}) reversing backward scattering
to forward one.
\begin{figure}
\begin{center}
\epsfxsize88mm \epsfbox{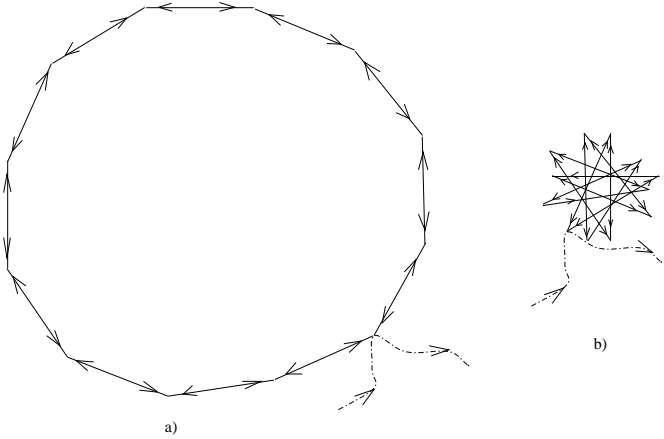}
\end{center}
\caption{a) Self- intersecting trajectory due to multiple scattering
  on impurities with small tilted angles ${\alpha}_i$ in $i-th$ act of
  scattering;(b)The same trajectory that is drawn in (a) 
  with the exception that each scattering is accompanied by
  BR, resulting in angles ${\alpha}_i + \pi$.In the both figures the
  magnitudes of velocity vectors are chosen to be the same.} 
\label{Fig.1}
\end{figure}
 The expression for the two particle impurity block $C_{\pi}({\bf
  q},\epsilon +{\epsilon}')$ in the particle--particle channel  due to
Umklapp scatterings with the particles' energies $\epsilon$ and
$\epsilon '$ is given as, 
\cite{nko,nfok,np}: 
\begin{eqnarray}
C_{\pi}({\bf q},\epsilon +{\epsilon}')&=& \frac{1}{2\pi\rho_{od}
  \tau_o}\Big\{\theta(-\epsilon {\epsilon}') +\nonumber\\
&&\frac{\theta(\epsilon {\epsilon}')}{(1-i\tau_o |\epsilon
  +{\epsilon}'| + \gamma \tau_o)^2 +\frac{2}{d} (ql)^2 - 1}\Big \},
\label{UCooperon}
\end{eqnarray}
where the  phenomenological parameter $\gamma$ is introduced to signify an
inelastic processes rate.
The $\pi$-Diffuson $D_{\pi}({\bf q}, \epsilon +{\epsilon}')$ is also
expressed by Eq.(\ref{UCooperon}) with exception that ${\bf q}$ will
now be the momenta difference of a particle and a hole with accuracy of
the nesting vector  ${\bf Q}$. Notice that the 'normal' Cooperon
and Diffuson blocks depend on the difference of the energies $\epsilon$ and
$\epsilon '$ instead of sum in Eq.(\ref{UCooperon}).

New diagrams (see, Fig.2) appear at half--filling due to BR which give
contributions to $R(\epsilon,\epsilon ')$ in addition to that coming
from normal scattering in a diffusive system. The study of the
diffusive regime assumes that the linear length $L$ of the system is
larger than the elastic mean free path $l$ and smaller than the
localization length, the latter of which is exponentially large. By 
considering the energy scales as the Thouless energy, $E_c = \hbar
D/L^2$ with the diffusion coefficient $D =\frac{v_F^2 \tau_o}{d}$ of $d$
dimensional system,and the average level spacing, $\Delta =
\frac{1}{\rho_{od} L^d}$, it is possible to see that  $\frac{E_c}{\Delta}
= \hbar \rho_{od} DL^{d-2} = \frac{\sigma}{(e^2/\hbar )}L^{d-2}=g$ is a 
dimensionless conductance. Since $g>1$ for the metallic case, the
diffusive system can be characterized by the condition  $\Delta \ll E_c
\ll \hbar/\tau_o$.
By summing up contributions of the diagrams in Fig.2 the correlator
$R(\epsilon, {\epsilon}')$ is expressed as,
\begin{eqnarray}
&&R(\epsilon, {\epsilon}')= \frac{(s \Delta \tau_o)^2}{\beta {\pi}^2} Re
\sum_{\bf q}\Big\{\frac{\theta(-\epsilon
  {\epsilon}')}{\tau_o^2(-i|\epsilon -{\epsilon}'| + Dq^2 +\gamma)^2}
-\nonumber\\ 
&&-\frac{\theta(\epsilon {\epsilon}')}{[(1-i\tau_o|\epsilon +
  {\epsilon}'| +\gamma \tau_o)^2 + \frac{2}{d} (ql)^2 -1]^2}\Big\},
\label{cor}
\end{eqnarray}
where  $q^2 = \sum_{\alpha =1}^d q_{\alpha}^2$  and
$q_{\alpha}=(2\pi/aN_{\alpha})n_{\alpha}$ 
with $(-N_{\alpha}/2)<n_{\alpha}\le (N_{\alpha}/2)$, and $\beta$ is
the Dyson index classifying the orthogonal, unitary and symplectic
ensembles with $\beta =$ 1, 2, and 4, respectively, \cite{mehta}. Far
from half--filling where BR are suppressed, only the first term  in the
bracket of Eq.(\ref{cor}) contributes to $R$,\cite{as}. 
\begin{figure}
\begin{center}
\epsfxsize88mm \epsfbox{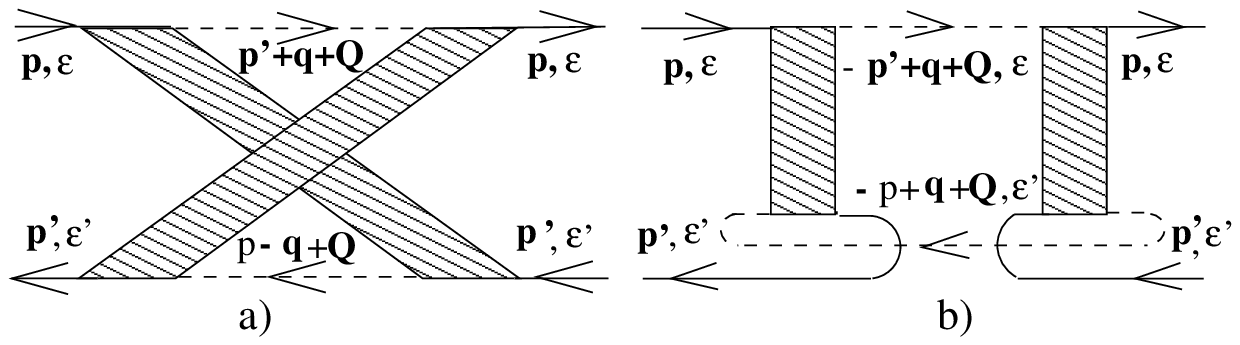}
\end{center}
\caption{First order additional corrections to the DoS correlator $R(\epsilon,
  {\epsilon}')$ in the metallic
 regime due to BR from a) $\pi$--Diffuson and b) $\pi$--Cooperon blocks.} 
\label{Fig.2}
\end{figure}
In the ergodic regime, $|\epsilon \pm {\epsilon}'|\ll E_c$, only
the  ${\bf q}=0$ term needs to be retained in
the summation over ${\bf q}$ in Eq.(\ref{cor}). So,
\begin{equation}
R(\epsilon, {\epsilon}')= -\frac{(s\Delta)^2}{\beta {\pi}^2}
Re\Big\{\frac{\theta(-\epsilon {\epsilon}')}{(\epsilon
  -{\epsilon}'+i\gamma)^2} - \frac{\theta(\epsilon
  {\epsilon}')}{4(\epsilon +{\epsilon}' +i\gamma)^2}\Big\},
\end{equation}
which shows that two levels on the opposite sides of the Fermi surface
with energy difference $|\epsilon - {\epsilon}'|>\gamma$ repel each
other and they attract at energy difference of $|\epsilon - {\epsilon}'|<
\gamma$. On the other hand, two levels with energies $|\epsilon +{\epsilon}'|>
\gamma$ on the same side of the Fermi surface attract each other and
they repel for the energies $|\epsilon +{\epsilon}'|<
\gamma$. Furthermore, attraction of two levels on the Fermi surface
($\epsilon = {\epsilon}' = 0$) weakens with approaching half--filling
and the correlator $R(0,0)$ reachs its $3/4$ value at
half--filling. Notice that far from half--filling the levels lying
only on 
the opposite sides of the Fermi surface interact with each
other. Additional interaction of the levels on the same side of the
Fermi surface appears due to BR at half--filling.

The correlator of two levels centered at ${\epsilon}_o$ and
${\epsilon}'_o$ and averaged in an energy interval of $E \le W$, where
$W = 2d t$ is the band width, can be obtained by integrating
$R(\epsilon,{\epsilon}')$ given by Eq.(\ref{cor}) over the energy
interval $E$:
\begin{eqnarray}
&&\langle \delta \rho_{\epsilon_o}(E) \delta \rho_{{\epsilon}'_o}(E)
\rangle =\int_{\epsilon_o-E/2}^{\epsilon_o+E/2} 
d\epsilon \int_{\epsilon_o'-E/2}^{\epsilon_o'+E/2} d{\epsilon}'
R(\epsilon,{\epsilon}') \nonumber\\
&&=\frac{s^2}{\beta \pi^2 \rho_{od}^2} Re \Big \{ \ln
\frac{{\gamma}^2[(\epsilon_o - \epsilon_o')^2 -
(E+i\gamma)^2]}{[\epsilon_o^2 - (E/2 +i\gamma)^2][{\epsilon_o'}^2 
  - (E/2 +i\gamma)^2]}- \nonumber\\
\label{dos1}
&&-\frac{1}{4}\ln \frac{{\gamma}^2[(\epsilon_o + \epsilon_o')^2
  - (E+i\gamma)^2]}{[\epsilon_o^2 - (E/2 +i\gamma)^2][{\epsilon_o'}^2
  - (E/2 +i\gamma)^2]}\Big \};\nonumber\\
&&\qquad \qquad |\epsilon_o|,|\epsilon_o'|\le E/2\\
&&=\frac{s^2}{\beta \pi^2 \rho_{od}^2} Re \ln \Big[ 1 -
\frac{E^2}{(|\epsilon_o|+|\epsilon_o'| 
  +i\gamma )^2}\Big]\nonumber\\ 
&&\times \{\theta(-\epsilon_o \epsilon_o') -
\frac{1}{4}\theta(\epsilon_o \epsilon_o')\};\quad
|\epsilon_o|,|\epsilon_o'|\ge E/2. 
\label{dos2}
\end{eqnarray}
Far from half--filling the second contribution in the brackets of
Eqs.(\ref{dos1}) and (\ref{dos2}) vanishes. This case corresponds to
the continuum model,\cite{as}. However, the fact that
interactions of levels lying on opposite sides of the Fermi
surface do give contribution to $ \langle \delta \rho_{\epsilon_o}(E)
\delta \rho_{{\epsilon}'_o}(E) \rangle$ has not been 
taken into account in \cite{as}. As it is seen from
Eqs.(\ref{dos1})-(\ref{dos2}) the  
two--level correlation function strongly depends on center of energy
strip $E$ even a correlation is considered in the same energy interval
when $\epsilon_o = \epsilon_o'$. A logarithmical energy dependence
of the variance $\langle [\delta \rho_{\epsilon_o}(E)]^2 \rangle$
takes place for a strip centered around the Fermi level:
\begin{eqnarray}
\langle [\delta \rho_{\epsilon_o =0}(E)]^2 \rangle &=&\frac{2s^2 
(1-f)}{\beta {\pi}^2\rho_{od}^2}\ln \frac{E}{\gamma};\quad E/2 <
\gamma < E,\nonumber\\ 
&=&-\frac{2s^2(1-f)}{\beta \pi^2 \rho_{od}^2}\ln
\frac{E}{4\gamma};\quad\gamma <  E/2,
\label{var}
\end{eqnarray}
where  $f$ is the parameter characterizing the BR: $f =1/4$ at half--filling and $f =0$ far from 
commensurate points. According to Eq.(\ref{var}) the Dyson repulsion
of levels for energies  $E/2 < \gamma < E$ turns to attraction of levels for 
large energy distances $E > 2\gamma $.

For the diffusive limit, when $E \gg E_c$, summing over ${\bf q}$ in
Eq. (\ref{cor}) can be replaced by integration. As a result we get the
following expressions for the DoS variance $ \langle [\delta
\rho_{\epsilon_o =0}(E)]^2 \rangle$ : 
\begin{eqnarray}
\langle [\delta \rho_{\epsilon_o =0}(E)]^2 \rangle _{dif}
&=&\frac{(\sqrt
  2-1)(1-f)s^2}{6{\pi}^3\beta
\rho_{o3}^2}\big(\frac{E}{E_c}\big)^{3/2};\quad d=3\nonumber\\ 
&=&-\frac{(1-f)s^2}{4{\pi}^3\beta
  \rho_{o2}^2}(E\tau_o)\big(\frac{E}{E_c}\big);\quad d=2.
\label{dif}
\end{eqnarray}
In the $d=2$ case linear contributions in $E$ to $ \langle [\delta
\rho_{\epsilon_o =0}(E)]^2 \rangle$ in Eq.(\ref{dif}) are completely
cancelled and the fluctuations are not as strong as
in $3D$ systems. This seems to be connected with the localized character of
levels in $2D$ systems.
\begin{figure}
\begin{center}
\epsfxsize88mm \epsfbox{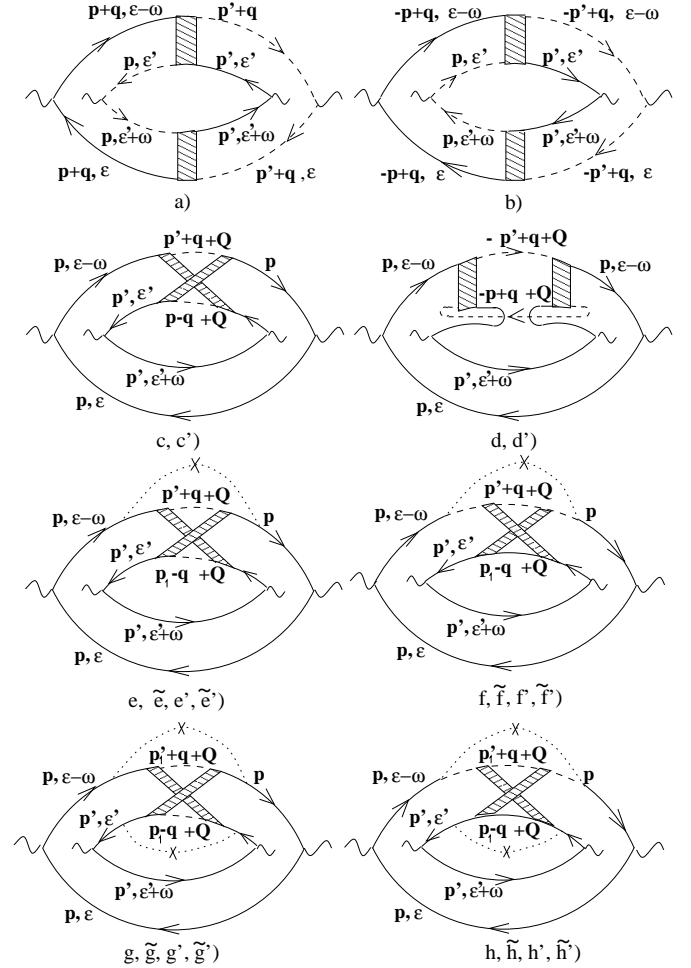}
\end{center}
\caption{The diagrams which contribute to the conductivity
  variance due to BR. The diagrams denoted by primes differ
  from the presented ones through the direction of the electron
  lines. ($\tilde{e}$),($\tilde{f}$) are symmetric to (e),(f)
  with respect to the single
  impurity line; and ($\tilde{g}$), ($\tilde{h}$) are obtained from
  (g),(h) by interchanging
  the straight and dashed lines under single impurity lines. Diagrams,
  similar to (e)-($\tilde{h'}$) exist also in the Cooper channel which
  are produced from ($d,d'$).}
\label{Fig.3}
\end{figure}
Far from half--filling when the Fermi surface is approximately spheric
the variance of fluctuations in static conductivity is designated by 
the diagrams given, e.g. in Fig. 4 of Ref.\cite{as}. These diagrams
have been calculated also in 
\cite{altshuler,ls,akl} for continuum and isotropic
systems. Other contributions to the conductivity variance exist in
the lattice model under consideration at the commensurate points due
to BR which come from the diagrams given in Fig.\ref{Fig.3}. 
The total contributions to the conductance
fluctuations due to Normal and Umklapp scatterings on impurities are:
\begin{equation}
\langle \delta G_{\alpha \beta} \delta G_{\gamma \mu} \rangle = G_D^2
\{\delta_{\alpha \gamma} \delta_{\beta \mu} + \delta_{\alpha \mu}
\delta_{\beta \gamma} \} + G_{\rho}^2 \delta_{\alpha \beta}
\delta_{\gamma \mu}.
\end{equation}
Here we followed the notations in \cite{as}, where the
contributions from the diffusion coefficient and the DoS fluctuations
to the conductance variance $ \langle \delta G_{\alpha \beta} \delta
G_{\gamma \mu} \rangle$ were denoted by the temperature
dependent coefficients $G_D^2$ and $G_{\rho}^2$,
respectively. The expressions for $G_D^2$ and $G_{\rho}^2$ can be
presented as:
\begin{eqnarray}
G_D^2 &=&\frac{s^2}{\beta}\big(\frac{e^2}{\hbar}\frac{E_c}{\pi}\big)^2\int
\frac{d\epsilon}{2T}f\big(\frac{\epsilon}{2T}\big)\sum_{\bf
  q}\big\{\frac{1}{|D q^2 - i\epsilon |^2} +\nonumber\\
&+& \frac{\tau_o^2}{|(1-i\epsilon \tau_o)^2 +\frac{2}{d}(ql)^2 -1|^2}\big\}
\label{G_D}
\end{eqnarray}
and
\begin{equation}
G_{\rho}^2 =
\frac{s^2}{\beta}\big(\frac{e^2}{\hbar}\frac{E_c}{\pi}\big)^2 \int
\frac{d\epsilon}{2T}f\big(\frac{\epsilon}{2T}\big) Re \sum_{\bf
  q}\frac{1}{(D q^2 - i\epsilon )^2},
\label{G_rho}
\end{equation} 
where $f(x)= \frac{x \coth x -1}{{\sinh}^2x}$.
The second term in the brackets in Eq.(\ref{G_D}) comes
from the diagrams given in Fig.\ref{Fig.3} due to
$\pi$--scatterings. However BR give no contribution to  
$G^2_{\rho}$. According to Thouless picture \cite{thouless},only
one--electron states lying in an interval of $E_c$ centered on the
Fermi 
level give contributions to the conductivity. Contributions to
$G^2_{\rho}$ from the levels correlated on the same side of the Fermi
level seem to cancel each other.

At small temperatures $T\ll E_c$ the coefficients $G_D^2$ and
$G_{\rho}^2$ do not depend on temperature:
\begin{equation}
G_{\rho}^2 = (s^2/\beta)(e^2/{\pi}^3\hbar)^2 b_d \quad
\textrm{and}\quad G_D^2=(1+f)G_{\rho}^2         
\label{T=0}
\end{equation}
where $b_d$ is a constant, which depends on the system dimension.
In the case when $T\gg E_c$ the values of $G_D^2$ and $G_{\rho}^2$ strongly
differ from each other and depend on temperature:
\begin{equation}
G_{\rho}^2= (s^2/\beta) (e^2/{2\pi \hbar})^2 a_d (E_c/T)^{(4-d)/2},
\label{G_rhoT}
\end{equation}
and,
\begin{eqnarray}
&G_D^2&= \frac{1}{2} (1+f)G_{\rho};\quad d= 3\nonumber\\
&=&(1+f)(s^2/\beta) (e^2/{2\pi \hbar})^2 \frac{E_c}{T}\ln \frac{T}{max
  \{E_c, \gamma \}};\quad d=2,
\label{G_DT}
\end{eqnarray} 
where $a_d$ is some coefficient, \cite{as}.
As can be seen from Eqs.(\ref{G_rhoT}) and (\ref{G_DT}) the main
contribution to the conductance variance comes from the fluctuations
of the diffusion coefficient, which are intensified at half--filling.

Multiplication of the variance by the additional prefactor $(1+f)$
means that Umklapp scatterings  of electron on impurities change the
distribution function of $G$.
In the language of the random matrix theory, an insulating phase of a
disordered system can be prescribed by an ensemble of $N \times N$ diagonal
matrices with random elements. Existence of  off- diagonal
terms in $N \times N$ matrices due to overlapping states of different energy 
in diffusive systems  transform the distribution function from Poisson
function to Wigner- Dyson one. Umklapp scatterings on impurities
give additional contributions to the off- diagonal matrix
elements. Therefore, the change in the distribution function due to Umklapp
scattering is reasonable. 
Two-level correlations are sensitive to whether the levels
attract or repel each other and to the relative position of these
levels, either on the
same side or on the opposite sides of the Fermi
surface. However, the conductance fluctuations seem not to be
sensitive to the character of level interactions and the level
positions; both attraction and repulsion give similar contributions to
$\langle \delta G_{\alpha \beta} \delta G_{\gamma \mu} \rangle$. This
fact seems to be the reason why Eqs.(\ref{var})- (\ref{dif}) for
$\langle [\delta \rho_{\epsilon_o}(E)]^2 \rangle $ and
Eqs.(\ref{T=0}),(\ref{G_rhoT}),(\ref{G_DT}) contain the
factor $f$ coming from BR in different way.

Recently fabricated $C_{60}$--based novel field--effect devices
allow one to control the band filling by changing the  gate potential.
Half--filling is reached for 3 electrons doping per $C_{60}$ molecule,
\cite{batlogg}. The possibility of doping the fullerites by substitutional
impurities, while preserving the periodicity of the Bravais lattice,
will allow the observation of the commensurability effects on the
mesoscopic fluctuations at half--filling in these devices.

\end{document}